\documentclass[%
reprint,
superscriptaddress,
 amsmath,amssymb,
 aps,
pra,
]{revtex4-1}

\usepackage{graphicx}
\usepackage{dcolumn}
\usepackage{bm}

\def\be{\begin{eqnarray}}   
\def\ee{\end{eqnarray}}

\newcommand{\affA}{
Center for Computational Sciences, University of Tsukuba, Tsukuba 305-8577, Japan
}
\newcommand{\affB}{
Max Planck Institute for the Structure and Dynamics of Matter and Center for Free-Electron Laser Science, Luruper Chaussee 149, 22761 Hamburg, Germany
}
\newcommand{\affC}{
Center for Computational Quantum Physics (CCQ), The Flatiron Institute, 162 Fifth avenue, New York NY 10010
}
\newcommand{\affD}{
Nano-Bio Spectroscopy Group, Departamento de Fisica de Materiales, Universidad del Pa\'is Vasco UPV/EHU, 20018 San Sebasti\'an, Spain
}

\begin{document}

\preprint{APS/123-QED}

\title{Exact exchange-correlation potential of effectively interacting Kohn-Sham systems}

\author{Shunsuke~A.~Sato}\email{ssato@ccs.tsukuba.ac.jp}\affiliation{\affA}\affiliation{\affB}
\author{Angel~Rubio}\affiliation{\affB}\affiliation{\affC}\affiliation{\affD}

\date{\today}

\begin{abstract}
  Aiming to combine density functional theory (DFT) and wavefunction theory,
  we study a mapping from the many-body interacting system to
  an effectively-interacting Kohn-Sham system instead of 
  a non-interacting Kohn-Sham system.
  Because a ground state of effectively-interacting systems
  requires having a solution for the correlated many-body wavefunctions,
  this provides a natural framework to many-body wavefunction theories 
  such as the configuration interaction
  and the coupled cluster method in the formal theoretical framework 
  of DFT.
  Employing simple one-dimensional two-electron systems -- namely, the one-dimensional 
  helium atom, hydrogen molecule and heteronuclear diatomic molecule --
  we investigate properties of many-body wavefunctions and
  exact exchange-correlation potentials of effectively-interacting
  Kohn-Sham systems. As a result, we find that
  the asymptotic behavior of the exact exchange-correlation potential
  can be controlled by optimizing that of the effective interaction.
  Furthermore, the typical features of the exact non-interacting Kohn-Sham system,
  namely a spiky feature and a step feature
  in the exchange-correlation potential
  for the molecular dissociation limit can be suppressed by a proper choice
  of the effective interaction.
  These findings open a possibility to construct numerically robust
  and efficient exchange-correlation potentials and functionals
  based on the effectively-interacting Kohn-Sham scheme.
\end{abstract}

\maketitle


\section{Introduction \label{sec:intro}}

Density functional theory (DFT) is one of the most successful
approaches to describe the ground state properties of electronic systems
\cite{PhysRev.136.B864,PhysRev.140.A1133}.
A strong point of DFT is its computational feasibility
and often offering the best compromise of accuracy and computational costs.
The feasible computational cost of DFT calculations
can be achieved by the mapping from a fully-interacting problem
to a non-interacting problem based on the Hohenberg-Kohn theorems.
In this mapping, all the complexities of the many-body problem
are absorbed in the unknown exchange-correlation functional.
Therefore, the accuracy of the DFT calculations essentially depends
on the approximation of the exchange-correlation functionals.
In the past decades, various exchange-correlation functionals have been developed
to realize accurate description of the electronic ground state
such as local density approximations (LDA) \cite{PhysRevB.23.5048,PhysRevB.45.13244},
generalized gradient approximations (GGA) \cite{PhysRevA.49.2421,PhysRevLett.77.3865},
meta-GGA \cite{PhysRevLett.91.146401,doi:10.1063/1.476577,PhysRevLett.115.036402},
and hybrid functions \cite{doi:10.1063/1.464913,doi:10.1063/1.472933,doi:10.1063/1.1564060}.
Furthermore, detailed studies clarified
several exact properties of the exact exchange-correlation functional and potential
such as asymptotic behavior
of the potential in Coulombic systems
\cite{PhysRevA.30.2745,PhysRevA.29.2322,PhysRevA.49.2421},
and spiky features in the molecular dissociation
\cite{PhysRevA.40.4190,Gritsenko1997,doi:10.1063/1.3271392,PhysRevB.93.155146,Burke2006exactconditions}.
However, the systematic improvement of the exchange-correlation functionals and potentials
is still a non-trivial task due to
highly nonlinear and nonlocal natures of the density functional
\cite{doi:10.1063/1.1390175,Medvedev49}.

In contrast to DFT, wavefunction theories \cite{szabo1989book} such as the configuration
interaction and the coupled-cluster method
offer a formal possibility to straightforwardly improve 
the accuracy up to the exact solution
by increasing the size of a search space
although the required computational costs can easily become infeasible.
Furthermore, in different fields, 
various methods have been developed for accurate description of electronic
structure such as the \textit{GW} method
\cite{Aryasetiawan_1998,AULBUR20001,RevModPhys.74.601,vanSetten2013} and
the quantum Monte Carlo method
\cite{RevModPhys.73.33,PhysRevB.84.245117,PhysRevB.98.075122,PhysRevLett.120.025701}.
However, such accurate approaches require huge computational costs
and they become infeasible for large systems.

Because DFT and wavefunction theory are based on different
characteristics,
the two approaches often have different points of strength and weakness.
For example, DFT with conventional approximations tends to well
capture the \textit{dynamical} correlation effect \cite{doi:10.1063/1.474864},
which requires many Slater determinants for accurate description of many-body
wavefunctions, while DFT 
suffers from dramatical failures in describing 
the \textit{static} correlation \cite{Cohen792,PhysRevLett.102.066403,doi:10.1063/1.3271392,Ess2011},
which requires a few but more than one Slater determinant
for accurate description of many-body wavefunctions.
On the other hand, the wavefunction theory
can naturally capture the static correlation effect through the configuration interaction.
Combining the two approaches, one may be able to realize an overall accurate
theoretical descriptions for quantum many-body systems.
Hybrid functional \cite{doi:10.1063/1.464913,doi:10.1063/1.472933} 
is one of successful examples,
and it includes a part of nonlocal Fock-like exchange interaction based on 
the wavefunction theory.
Importantly, note that hybrid functional is beyond the conventional theoretical framework
of DFT but it is based on the generalized Kohn-Sham scheme \cite{PhysRevB.53.3764},
where interacting model systems are introduced to take into account a part
of electron-electron interaction but the systems are still represented
by a single Slater determinant.
Thus, the generalized Kohn-Sham systems are still described by fully-uncorrelated
wavefunctions.
Another successful example is a combination of 
the configuration interaction method and DFT 
\cite{GRIMME1996128,Borowski1998,doi:10.1063/1.479866,GRAFENSTEIN1998593,FILATOV1998689,doi:10.1063/1.4804607}
where the Kohn-Sham orbitals based on DFT are used to construct the configuration
interaction approach.
Since the Kohn-Sham orbitals and their orbital energies 
may take into account substantial amount of the dynamical correlation effect, 
the configuration interaction based on the DFT approach drastically 
improves the description.

In this work, we explore yet another possibility to accurately and efficiently 
combine DFT and wavefunction
theory, introducing a mapping between a fully-interacting many-body system
and an \textit{effectively-interacting} Kohn-Sham system instead of
the non-interacting Kohn-Sham system.
Fromager \textit{et al.} have achieved the connection between a fully-interacting system
and a fictitious interacting system in terms of the range separation in
multiconfigurational density functional theory \cite{doi:10.1063/1.2566459}.
Effectively-interacting systems are not generally described by
a single Slater determinant but requires a correlated wavefunction.
Therefore, the wavefunction theory may be naturally introduced within 
the formal theoretical framework of DFT. By optimally choosing the effective interaction,
the electronic correlation may be efficiently described by
the combination of DFT and wavefunction theory.
To explore more in detail the properties of an effectively interacting Kohn-Sham systems,
we investigate how the exact exchange-correlation potential looks like for
that correlated reference Kohn-Sham system, employing one-dimensional
two-electron systems.

The paper is organized as follows: In Sec.~\ref{sec:methods}
we first introduce a mapping between a fully-interacting system and
an effectively-interacting Kohn-Sham system.
Then, we describe effective interactions that are used in this work.
In Sec.~\ref{sec:numerics}, numerical methods to compute exact exchange-correlation
potentials of effectively-interacting Kohn-Sham systems are described.
In Sec.~\ref{sec:results},
we investigate exact exchange-correlation potentials
and properties of effectively-interacting Kohn-Sham systems
for one-dimensional helium atom, hydrogen molecules
and heteronuclear diatomic molecule.
Finally, our findings are summarized in Sec.~\ref{sec:summary}.

\section{Methods \label{sec:methods}}

We introduce a mapping from a fully interacting many-body system to
an effectively-interacting Kohn-Sham system. For this purpose, we first
consider a fully-interacting $N$-electron system.
The ground state of
the electronic system is described by the following Schr\"odinger equation
\be
\hat H \Psi(\bm r_1,\cdots, \bm r_N) = E_{gs} \Psi(\bm r_1,\cdots,\bm r_N),
\label{eq:exact-se}
\ee
with the non-relativistic many-body Hamiltonian given by
(atomic units used unless stated otherwise)
\be
\hat H = \sum_{i=1}^N \left [
  -\frac{1}{2}\bm{\nabla}^2_i +v_{ext}(\bm{r}_i) \right ]
+\frac{1}{2}\sum_{i\neq j} W(\bm r_i-\bm r_j),
\label{eq:many-body-ham}
\ee
where $v_{ext}(\bm r)$ is a one-body external potential and $W(\bm r)$ is
the electron-electron interaction.

Then, we introduce an effectively-interacting Kohn-Sham system that
satisfies the following interacting Kohn-Sham equation
\be
\hat H_{KS} \Phi_{KS}(\bm r_1,\cdots,\bm r_N) = E_{KS} \Phi_{KS}(\bm r_1,\cdots,\bm r_N)
\label{eq:ks-eq}
\ee
with the interacting Kohn-Sham Hamiltonian
\be
\hat H_{KS} = \sum_{i=1}^N \left [
  -\frac{1}{2}\bm \nabla^2_i +v_{KS}(\bm r_i) \right ]
+\frac{1}{2}\sum_{i\neq j} W_{eff}(\bm r_i-\bm r_j),\nonumber \\
\label{eq:ks-ham}
\ee
where $v_{KS}(\bm r)$ is the one-body Kohn-Sham potential, and $W_{eff}(\bm r)$
is an arbitrary effective interaction.
Here, the Kohn-Sham potential $v_{KS}(\bm r)$ is introduced
such that the ground-state density of the fully-interacting system
is reproduced by that of the Kohn-Sham system
\be
\rho(\bm r)&=&N\int d\bm r_2\cdots d\bm r_N |\Psi(\bm r,\bm r_2,\cdots,\bm r_N)|^2 \nonumber \\
&=&N\int d\bm r_2\cdots d\bm r_N |\Phi_{KS}(\bm r,\bm r_2,\cdots,\bm r_N)|^2.
\ee
Note that here the Kohn-Sham potential $v_{KS}(\bm r)$ is not uniquely constructed
since a constant shift of the potential does not affect the ground state density.
Furthermore, the reconstruction of the corresponding energy functional or
exchange-correlation functional is not trivial.
Levy and Zahariev proposed a simple way to evaluate the exact
interacting ground state energy as a sum of orbital energies based on
the arbitrary constant term in the Kohn-Sham potential
\cite{PhysRevLett.113.113002}.
In this work, we extend this idea to interacting Kohn-Sham systems
and set the arbitrary constant in the Kohn-Sham potential $v_{KS}(\bm r)$
such that the ground state energy of the Kohn-Sham system $E_{KS}$
is identical to that of the fully-interacting system $E_{gs}$.

The Hohenberg-Kohn theorems offer one-to-one correspondence
between the ground state density $\rho(\bm r)$ and
the one-body external potential $v_{ext}(\bm r)$ once the interaction $W(\bm r)$
is given \cite{PhysRev.136.B864}.
Because the Hohenberg-Kohn theorems are not limited to the Coulomb interaction
but generally applicable to arbitrary interactions,
there is one-to-one correspondence between the one-body potential $v_{KS}(\bm r)$
and the corresponding ground state electron density once the effective interaction
$W_{eff}(\bm r)$ is defined.
Thus, once both interactions, $W(\bm r)$ and $W_{eff}(\bm r)$, are given,
there is a one-to-one correspondence between $v_{ext}(\bm r)$ and $v_{KS}(\bm r)$
through the common ground state density $\rho(\bm r)$,
resulting in a one-to-one correspondence between the fully-interacting many-body system
and the effectively-interacting Kohn-Sham system.
If the effective interaction is set to zero, $W_{eff}(\bm r)=0$,
this one-to-one mapping is reduced to the conventional Kohn-Sham mapping
between the interacting system and the corresponding
non-interacting Kohn-Sham system.

For later convenience, we decompose the Kohn-Sham potential $v_{KS}(\bm r)$
into the external potential $v_{ext}(\bm r)$,
the residual Hartree potential $v_{R-H}(\bm r)$
and the exchange-correlation potential $v_{xc}(\bm r)$.
Here, we define the residual Hartree potential as
\be
v_{R-H}(\bm r) = \int d\bm r' \Delta W_{res} (\bm r-\bm r')\rho(\bm r'),
 \nonumber \\
\label{eq:vh}
\ee
where $\Delta W_{res}(\bm r)$ is the residual interaction defined
as $\Delta W_{res}(\bm r) = W(\bm r)-W_{eff}(\bm r)$.
Note that, if the effective interaction is set to zero, $W_{eff}(\bm r)=0$,
the residual Hartree potential $v_{R-H}(\bm r)$
is reduced to the conventional Hartree potential,
$v_{H}(\bm r) = \int d\bm r' W(\bm r-\bm r')\rho(\bm r')$.
Furthermore, if $W_{eff}(\bm r)=W(\bm r)$,
the residual Hartree potential vanishes. 
Therefore, the residual Hartree potential, Eq.~(\ref{eq:vh}),
can be seen as a natural extension of the conventional Hartree potential.

Then, we define the exchange-correlation potential
as the rest of the Kohn-Sham potential
\be
v_{xc}(\bm r):= v_{KS}(\bm r)-v_{ext}(\bm r)-v_{R-H}(\bm r).
\label{eq:vxc}
\ee
The exchange-correlation potential in Eq.~(\ref{eq:vxc})
is reduced to the conventional exchange-correlation potential
of the non-interacting Kohn-Sham system if the effective interaction $W_{eff}(\bm r)$
is set to zero. Thus, Eq.~(\ref{eq:vxc}) can be also seen
as an extension of the conventional exchange-correlation potential.

In this work, we investigate effectively-interacting
Kohn-Sham systems and their exact exchange-correlation potentials
to explore a possibility to combine DFT and wavefunction
theory. To practically elucidate the effectively-interacting Kohn-Sham systems,
we consider the example of one-dimensional spin-$1/2$ two-electron systems.
As the electron-electron interaction, we employ the one-dimensional
soft Coulomb potential
\be
W(x) = \frac{1}{\sqrt{x^2 + \sigma^2}},
\ee
where $\sigma$ is a softening parameter, which is set to $0.5$~a.u.

As reference (interacting and non-interacting) Kohn-Sham systems,
we consider three kinds of systems in this work.
The first one is a \textit{non-interacting} Kohn-Sham system,
where the effective interaction $W_{eff}(x)$ is set to zero.
Note that this choice is nothing but the conventional non-interacting Kohn-Sham system
or standard DFT \cite{PhysRev.140.A1133}.
The second one is a \textit{$1/4$-interacting} system,
where the effective interaction is set to the quarter of
the bare soft Coulomb interaction, $W_{eff}(x)=W(x)/4$.
The third one is a \textit{long-range interacting} system,
where
$W_{eff}(x)=\mathrm{erf}(\sqrt{x^2+\sigma^2}/a_0)W(x)$ with
the Bohr radius $a_0$.
Thus, the short-range part of the Coulomb interaction is ignored
in the long-range interacting system.

\section{Numerical details \label{sec:numerics}}

Here, we describe numerical procedures to compute
the exact exchange-correlation potentials of the Kohn-Sham systems.
For the non-interacting two-particle Kohn-Sham system,
one can easily compute the exact Kohn-Sham potential as
\be
v_{KS}(x) = \frac{1}{2} \frac{1}{\sqrt{\rho(x)}}\frac{\partial^2}{\partial x^2}
\sqrt{\rho(x)}+E_{gs}.
\ee

For interacting Kohn-Sham systems, we employ the following
iterative scheme to obtain the exact Kohn-Sham potential
that reproduces the target ground-state density $\rho^{target}(x)$:
\begin{description}

\item[(i)] Start from an initial guess of the Kohn-Sham potential, $v^{(i=0)}_{KS}(x)$.
  In this work, we employ the external potential $v_{ext}(x)$
  as the initial guess.

\item[(ii)]
  Compute the ground-state density $\rho^{(i)}(x)$,
  solving the interacting Kohn-Sham equation, Eq.~(\ref{eq:ks-eq}),
  with the trial potential $v^{(i)}_{KS}(x)$.

\item[(iii)] Then, evaluate the deviation from the target density by
\be
r^{(i)}_{error} = \int dx \left| \rho^{(i)}(x)-\rho^{target}(x) \right|^2.
\ee

\item[(iv)] If the error $r^{(i)}_{error}$ is larger than a given threshold $\eta$,
the trial Kohn-Sham potential is updated by the following formula
\be
v^{(i+1)}_{KS}(x)=v^{(i)}_{KS}(x)+\alpha_i \frac{\rho^{(i)}(x)-\rho^{target}(x)}
{\rho^{(i)}(x)+\rho^{target}(x)+\epsilon},
\ee
where $\alpha_i$ is a mixing parameter, and $\epsilon$ is a small positive number.
In this work, we set $\epsilon$ to $10^{-8}$~a.u., and choose
the mixing parameter $\alpha_i$ such that
the error of the updated density $r^{(i+1)}_{error}$ in the next iteration
becomes smaller than that of the previous iteration $r^{(i)}_{error}$.
In practical calculations, we set the initial value of $\alpha_0=0.1$.
In the iterative procedure, we employ an acceptance-rejection procedure to determine
the value of $\alpha_i$. At each iteration, the initial guess of $\alpha_i$ is evaluated
as $\alpha_i = 1.1\alpha_{i-1}$. If the computed error $r^{(i+1)}_{error}$ is smaller
than the previous error $r^{(i)}_{error}$, the guess value of $\alpha_i$ is accepted.
If the computed error is not smaller than the previous error, the guess value of
$\alpha_i$ is rejected, and the new guess value is set
as a half of the previous guess value. This procedure is recursively repeated
until $r^{(i+1)}_{error}$ becomes smaller than $r^{(i)}_{error}$.

\item[(v)] Repeat the above iterative procedure until the error $r^{(i)}_{error}$
becomes smaller than a given threshold $\eta$.
In this work, we set the threshold $\eta$ to $10^{-9}$ a.u.

\end{description}

Note that similar iterative approaches to compute the exact Kohn-Sham potential have been
proposed \cite{PhysRevA.47.R1591,Nielsen2018}.
However, any stable approaches can be employed in this work 
because the size of the problems is small, and the accurate results can be obtained
with reasonable computational costs.

\section{Results \label{sec:results}}

\subsection{1D Helium atom \label{subsec:1d-he}}

First, we investigate the effectively-interacting Kohn-Sham systems
of the one-dimensional helium atom. To describe the helium atom, we employ the following
external potential
\be
v^{He}_{ext}(x)=-\frac{2}{\sqrt{x^2+\sigma^2}}.
\ee

\begin{figure}[htbp]
\centering
\includegraphics[width=0.9\columnwidth]{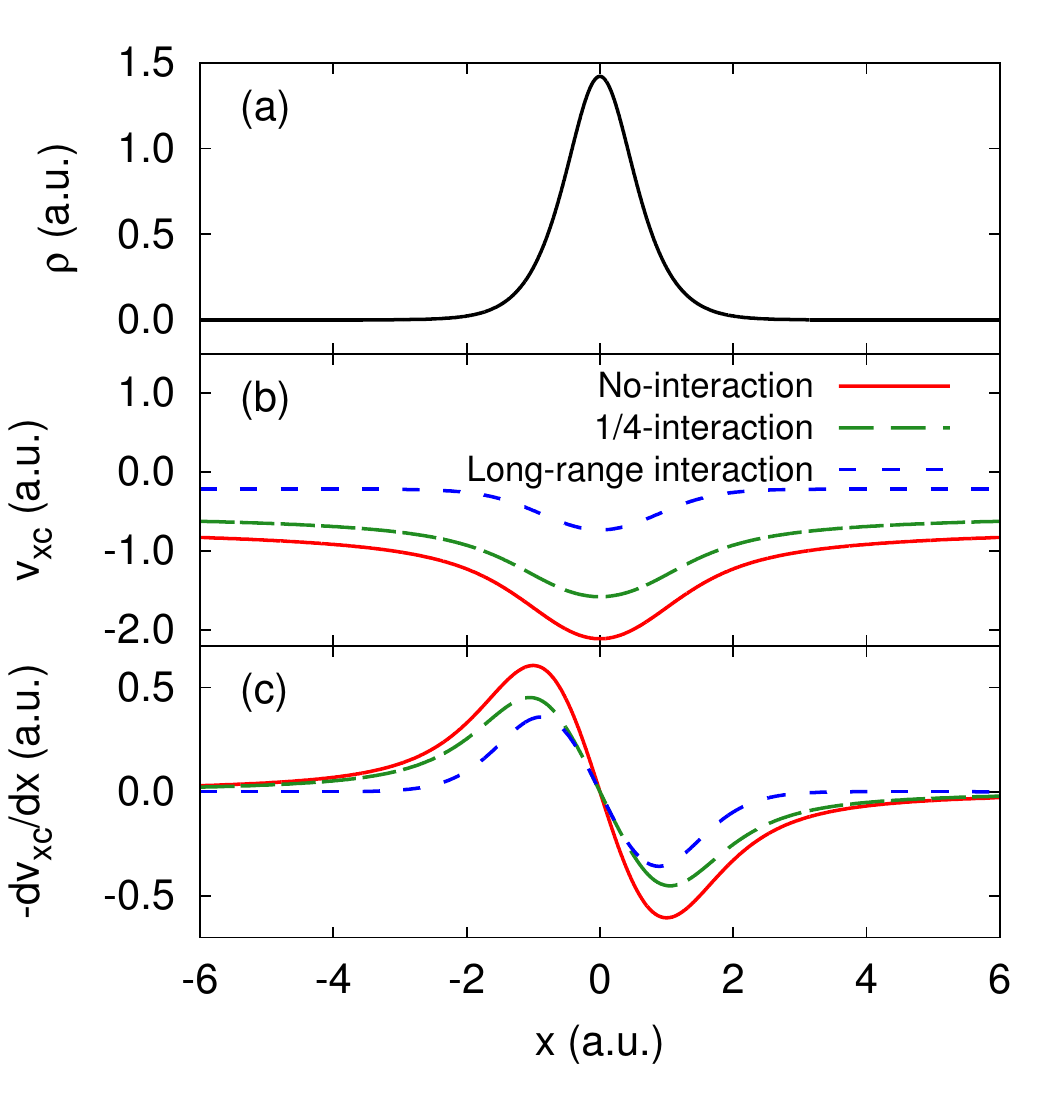}
\caption{\label{fig:he_vxc}
  (a) The exact ground-state density $\rho(x)$ of the one-dimensional
  helium atom. (b) The exact exchange-correlation potentials for
  the effectively-interacting Kohn-Sham systems: the red-solid line shows
  the results for the non-interacting system, $W_{eff}(x)=0$;
  the green-dashed line shows that for the $1/4$-interacting system, 
  $W_{eff}(x)=1/(4\sqrt{x^2+\sigma^2})$; 
  and the blue-dotted line shows the long-range interacting system, 
  $W_{eff}(x)=\mathrm{erf}(\sqrt{x^2+\sigma^2}/a_0)/\sqrt{x^2+\sigma^2}$. 
  (c) The force field $-v_{xc}(x)/dx$ of each exchange-correlation
  potential in the panel (b).
}
\end{figure}

Figure~\ref{fig:he_vxc}~(a) shows the exact ground-state density of the helium atom,
obtained by numerically solving the two-dimensional Schr\"odinger equation
with the conjugate gradient method.
Figure~\ref{fig:he_vxc}~(b) shows
the exact exchange-correlation potentials with respect to the three different interacting
Kohn-Sham systems. The different effective interactions
provide substantially different exchange-correlation potentials.
One may see that the long-range tail of the exact exchange-correlation potential
of the long-range interacting system (blue-dashed line) decays 
faster than those of the non-interacting and
the $1/4$-interacting Kohn-Sham systems.
To clearly compare the exchange-correlation potentials
of the different Kohn-Sham systems,
Figure~\ref{fig:he_vxc}~(c) shows
the exchange-correlation force field, $-dv_{xc}(x)/dx$.
One can clearly see that the exchange-correlation force field
of the long-range interacting
Kohn-Sham system shows much faster decay than
the other systems. Furthermore, the long-range interacting
system has the smallest force field over the whole spatial region.
These features indicate a possibility to control
the asymptotic behaviors of the exchange-correlation potential by
choosing a suitable effective interaction used to set the reference interacting
Kohn-Sham system.

\begin{figure}[htbp]
\centering
\includegraphics[width=0.9\columnwidth]{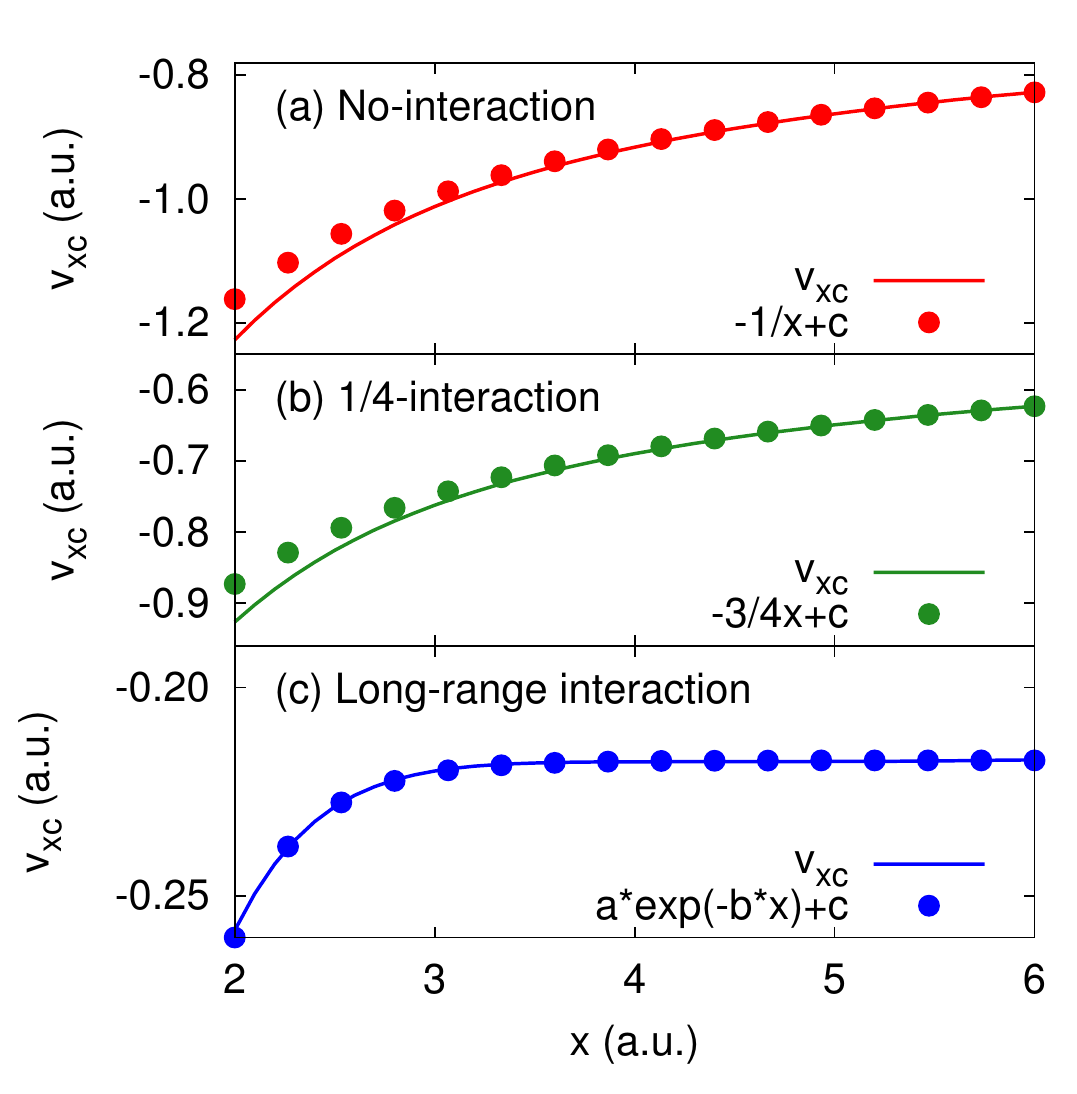}
\caption{\label{fig:he_vxc_ahympt}
  Asymptotic behaviors of the exact exchange-correlation
  potentials for (a) the non-interacting, (b) $1/4$-interacting
  and (c) the long-range-interacting Kohn-Sham systems.}
\end{figure}

Then, we investigate details of the asymptotic behavior
of the exchange-correlation potentials.
Figure~\ref{fig:he_vxc_ahympt} shows the long-range behavior
of the exact exchange-correlation potentials shown in Fig.~\ref{fig:he_vxc}~(b).
In Fig.~\ref{fig:he_vxc_ahympt}~(a), $v_{xc}(x)$ of the non-interacting system
is shown as the red line, while an analytic curve,
$-1/x+c$, is described by the red circles. Thus,
$v_{xc}(x)$ of the non-interacting system has $-1/x$ asymptotics.
This behavior is known as the asymptotic behavior of the exchange-correlation
potential of a Coulombic system \cite{PhysRevA.49.2421}.
In Fig.~\ref{fig:he_vxc_ahympt}~(b), $v_{xc}(x)$ of the $1/4$-interacting system
is shown as the green line, while the analytic curve,
$-\frac{3}{4x}+c$, is described by the green circles.
Therefore, $v_{xc}(x)$ of the $1/4$-interacting system
has $-\frac{3}{4x}$ asymptotics, which is different from
the conventional asymptotics of the exchange-correlation potential.
The $-\frac{3}{4x}$ asymptotics corresponds to the asymptotic behavior
of the residual interaction of the $1/4$-interacting system,
$\Delta W_{res}(x)=W(x)-W_{eff}(x)=3/(4\sqrt{x^2+\sigma^2})$.
Therefore, the asymptotic behavior of $v_{xc}(x)$
of an effectively-interacting Kohn-Sham system can be
characterized by that of the residual interaction.
In Fig.~\ref{fig:he_vxc_ahympt}~(c), $v_{xc}(x)$ of the long-range-interacting system
is shown as the blue line, while the analytical curve,
$a\times \exp[-bx]+c$, is described by the blue circles.
One sees that the asymptotic decay of $v_{xc}$ is slower than
that of the residual interaction,
$\Delta W_{res}(x)=W(x)-W_{eff}(x)=\mathrm{erfc}(\sqrt{x^2+\sigma^2}/a_0)/\sqrt{x^2+\sigma^2}$.
This fact can be understood by the asymptotic behavior of
the electron density: The residual interaction decays
so fast that the asymptotics of $v_{xc}(x)$ is dominated
by that of the electron density in order to
correctly remove the self-interaction error due to
the local density.
This feature may indicates that
the local density approximation for the exchange-correlation
functional may work well for the long-range interacting Kohn-Sham system
because the exchange-correlation potential vanishes once
the density vanishes and the nonlocality in the exchange-correlation
potential is expected to be milder than
that of the non-interacting Kohn-Sham system.

Due to effective interactions $W_{eff}(x)$,
the ground-state wavefunction of interacting Kohn-Sham Hamiltonian
in Eq.~(\ref{eq:ks-eq}) is not generally described by a single
Slater determinant, but it requires multi-Slater determinants
for the accurate description.
Therefore, in the interacting Kohn-Sham system,
the electronic correlation can be treated separately:
A part of the correlation can be treated as the explicit
correlation in the many-body Kohn-Sham wavefunction,
while the other part can be implicitly treated
in the exchange-correlation potential, $v_{xc}(x)$, or the corresponding
density functional $E_{xc}[\rho(x)]$.
This separation enables to combine DFT 
and wavefunction theory in order to efficiently
describe the electronic correlation. For example, the static correlation
may be efficiently treated by the multi-configuration interaction
as the explicit correlation in the correlated wavefunction, while
the dynamical correlation may be treated by DFT through
the exchange-correlation functional.
To explore such possibility,
we next investigate the explicit correlation
in the effectively-interacting Kohn-Sham wavefunctions.
For this purpose, we consider the eigendecomposition of the one-body
reduced density matrix
\be
\rho_{1RDM}(x,x')&=&2\int dx_2 \Psi_{KS}(x,x_2)\Psi^*_{KS}(x',x_2) \nonumber \\
&=&\sum_{i=1}n_i \phi_i(x)\phi^*_i(x'),
\ee
where eigenvectors $\phi_i(x)$ are known as natural orbitals \cite{PhysRev.97.1474},
and eigenvalues $n_i$ are seen as their occupations.
Here, we assume that
the occupation numbers are arranged in decreasing order
$n_i\ge n_{i+1}$.
Since we treat the spatial part of
the spin-singlet wavefunction, $\Psi_{KS}(x,x')$,
natural occupations are restricted as $0\le n_i \le 2$.

The occupation distribution deeply links to the number of
configurations that is required to accurately describe a correlated
electronic wavefunction
\cite{PhysRev.97.1474,szabo1989book}.
If only a small number of orbitals have substantial occupations
and the others have negligible occupations,
the correlated system can be described by
a small number of Slater determinants.
In contrast, if a larger number of orbitals have substantial
occupations, a larger number of configurations are required.
Electronic correlation in the first case is called
\textit{static} correlation, while that in the latter case
is called \textit{dynamical} correlation.

\begin{figure}[htbp]
\centering
\includegraphics[width=0.9\columnwidth]{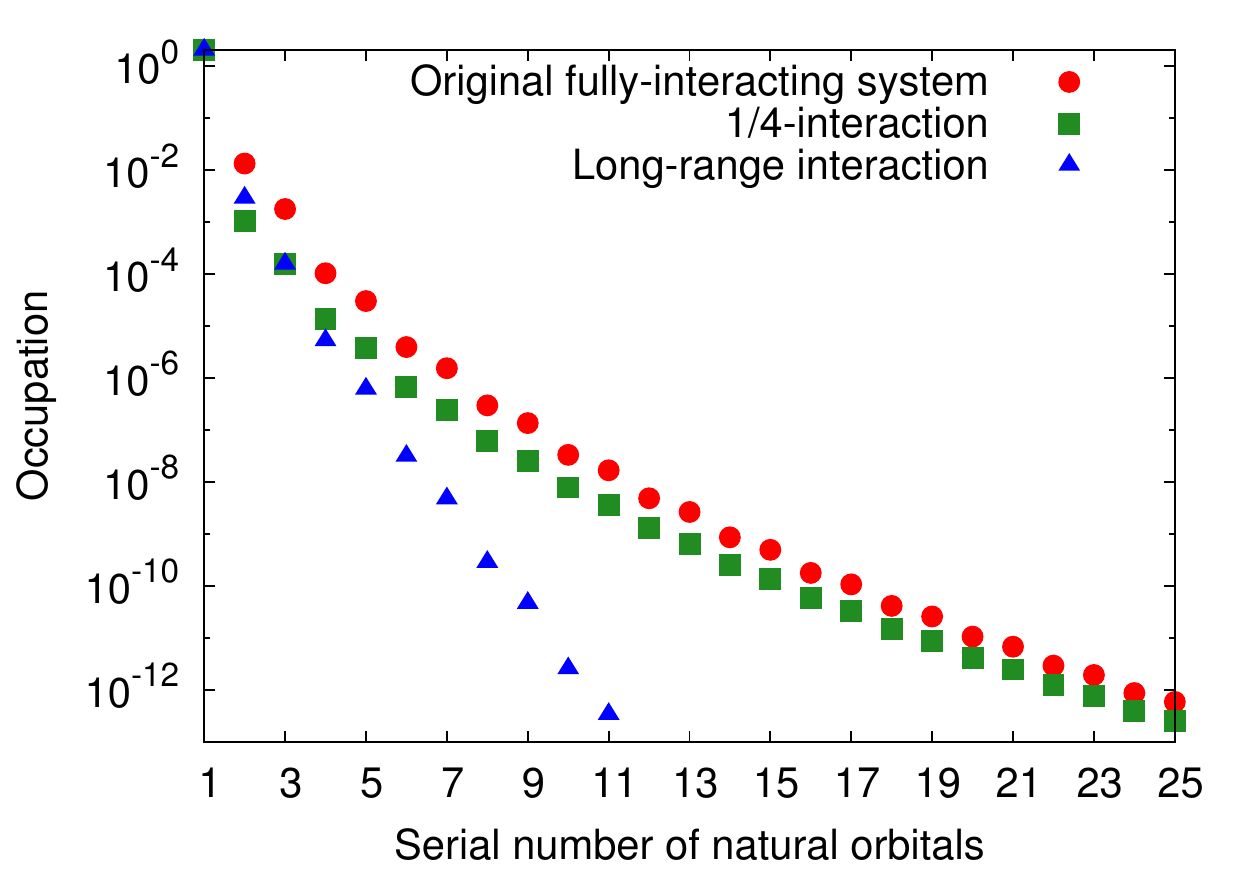}
\caption{\label{fig:natural_occ}
  Distribution of natural occupation over a few tens of
  natural orbitals. The results of the fully-interacting system
  (red circle), the $1/4$-interacting system (green square),
  and the long-range interacting system (blue triangle) are shown.
}
\end{figure}

Figure~\ref{fig:natural_occ} shows the distribution of the occupations $n_i$
of the correlated ground-state wavefunction of the one-dimensional
helium atom. The red circles show the occupation distribution
of the fully-interacting system, the green squares show that of
the $1/4$-interacting Kohn-Sham system,
and the blue triangles show that of the long-range interacting Kohn-Sham system.
As seen from the figure, the occupations of the higher natural orbitals
are significantly suppressed in the effectively-interacting Kohn-Sham systems,
compared with the fully-interacting problem.
Thus, a large part of the electronic correlation is transferred from
the explicit correlation in the many-body wavefunction to the exchange-correlation
functional.
Furthermore, in Fig.~\ref{fig:natural_occ},
the long-range interacting system shows the rapid decrease of the occupations
of the higher natural orbitals.
This fact indicates that, in the long-range interacting Kohn-Sham system,
the significant part of the dynamical correlation is transferred to
the exchange-correlation functional, while the static correlation
is treated as the explicit correlation in the reference correlated wavefunction.
Importantly, we note that the long-range interacting Kohn-Sham system has the weakest
exchange-correlation potential with the fastest asymptotic decay
among all the investigated Kohn-Sham systems (see Fig.~\ref{fig:he_vxc}).
Therefore, this fact clearly demonstrates that
a proper choice of the effective interaction
enables to efficiently decompose the electronic correlation
into the exchange-correlation functional part
and the explicit wavefunction correlation part, resulting in
an efficient description of the electronic correlation
based on a combination of DFT and wavefunction theory.

\subsection{1D H$_2$ molecule \label{subsec:1d-h2}}

Next, we investigate the effectively-interacting Kohn-Sham systems
of the one-dimensional hydrogen molecule.
To describe the hydrogen molecule,
we employ the following external potential
\be
v^{H_2}_{ext}(x)=-\frac{1}{\sqrt{(x-\frac{R}{2})^2+\sigma^2}}
-\frac{1}{\sqrt{(x+\frac{R}{2})^2+\sigma^2}}, \nonumber \\
\ee
where $R$ is the distance of the hydrogen atoms.
In this work, we set $R$ to $5$~a.u.

\begin{figure}[htbp]
\centering
\includegraphics[width=0.9\columnwidth]{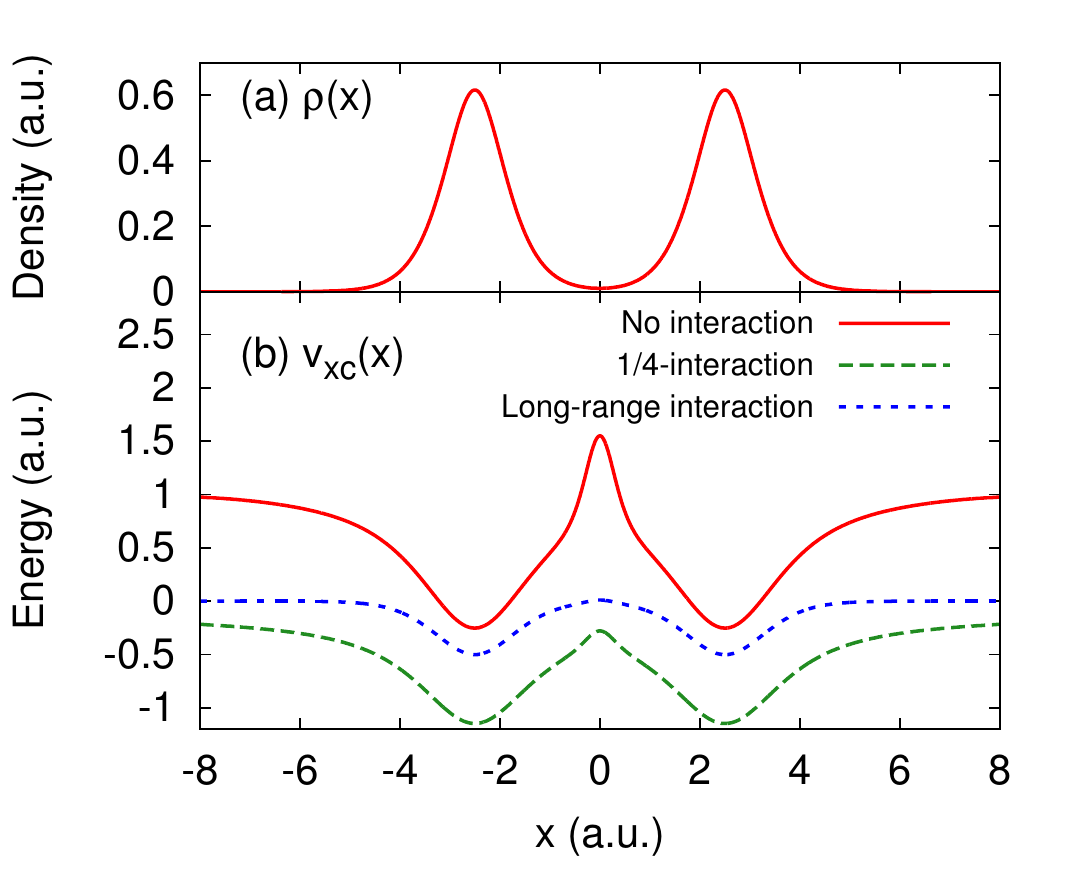}
\caption{\label{fig:vxc_h2}
  (a) The exact ground-state density $\rho(x)$ of the one-dimensional
  hydrogen molecule.
  (b) The exact exchange-correlation potentials for
  the effectively-interacting Kohn-Sham systems: the results for
  the non-interacting system (red-solid),
  the $1/4$-interacting system (green-dashed),
  and the long-range interacting system (blue-dotted) are shown.
}
\end{figure}

Figure~\ref{fig:vxc_h2}~(a) shows the exact ground state
electron density of the one-dimensional hydrogen molecule,
obtained by numerically solving the two-dimensional Schr\"odinger equation
with the conjugate gradient method.
At the center of the two hydrogen atoms,
the electron density becomes close to zero.
Figure~\ref{fig:vxc_h2}~(b) shows the exact exchange-correlation potentials
of the non-interacting and the effectively-interacting Kohn-Sham systems.
The exact exchange-correlation potential of the non-interacting system
(red-solid line) shows a spiky structure at the center.
This spiky structure was investigated in the previous works
for the strongly correlated system
\cite{doi:10.1063/1.3271392,PhysRevB.93.155146,PhysRevA.54.1957,PhysRevB.90.241107}.
One sees that the peak structure is strongly suppressed
in the $1/4$-interacting Kohn-Sham system (green-dashed line).
Furthermore, the exchange-correlation potential
of the long-range interacting Kohn-Sham system
(blue-dotted) shows very smooth feature around the central region.
This fact indicates that while the strong correlation effect
is encoded in the spiky structure
in the non-interacting Kohn-Sham system,
that of the effectively-interacting Kohn-Sham systems
is directly taken care through the explicit correlation
of the reference wavefunction.
Therefore, by properly choosing the effective interaction
of the Kohn-Sham system,
one can transfer the electronic correlation effect from
the exchange-correlation functional/potential
to the explicit correlation in the wavefunction in order to
reduce the complexity for developing accurate approximations
for the remaining exchange-correlation functional and potential.

\subsection{1D heteronuclear diatomic molecule \label{subsec:1d-hetero-molecule}}

Finally, we investigate the effectively-interacting Kohn-Sham systems of the one-dimensional
heteronuclear diatomic molecule.
In order to describe the heteronuclear diatomic molecule,
we employ the following external potential
\be
v^{HM}_{ext}(x)=-\frac{1-\delta}{\sqrt{(x-\frac{R}{2})^2+\sigma^2}}
-\frac{1+\delta}{\sqrt{(x+\frac{R}{2})^2+\sigma^2}}, \nonumber \\
\ee
where $R$ is the distance of the hydrogen atoms, and $\delta$ is the charge imbalance
between the two nuclei.
In this work, we set $R$ to $5$~a.u, and $\delta$ to $0.2$~a.u.

\begin{figure}[htbp]
\centering
\includegraphics[width=0.9\columnwidth]{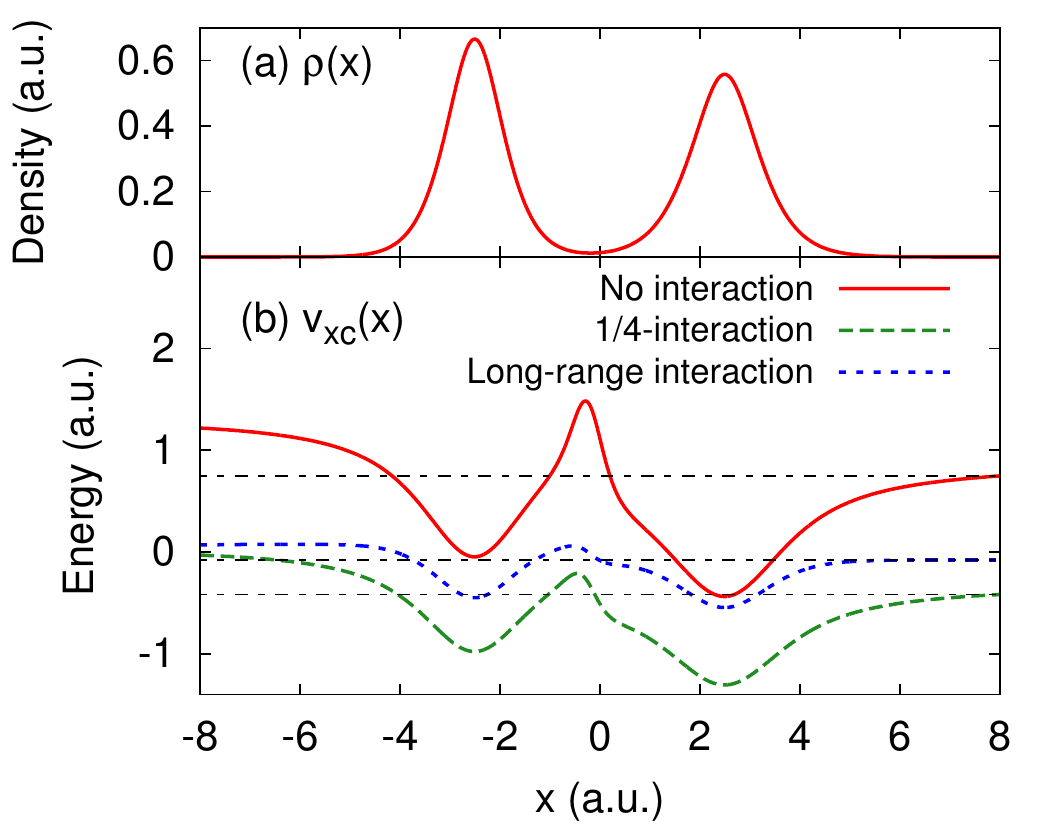}
\caption{\label{fig:vxc_hetero_molec}
  (a) The exact ground-state density $\rho(x)$ of the one-dimensional
  heterocuclear diatomic molecule.
  (b) The exact exchange-correlation potentials for
  the effectively-interacting Kohn-Sham systems: the results for
  the non-interacting system (red-solid),
  the $1/4$-interacting system (green-dashed),
  and the long-range interacting system (blue-dotted) are shown.
  horizontal black-dotted lines show values of each potential $v_{xc}(x)$
  evaluated at $x=8$~a.u.
}
\end{figure}

Figure~\ref{fig:vxc_hetero_molec}~(a) shows the exact ground state
electron density of the one-dimensional heteronuclear diatomic molecule,
obtained by numerically solving the two-dimensional Schr\"odinger equation
with the conjugate gradient method.
Reflecting the charge imbalance between two nuclei, the electron density $\rho(x)$
shows the asymmetric structure.
Figure~\ref{fig:vxc_hetero_molec}~(b) shows the exact exchange-correlation
potentials of the effectively-interacting Kohn-Sham systems.
The exchange-correlation potential of the non-interacting Kohn-Sham system
(red-solid line) asymptotically approaches to different values in the positive and negative
$x$ regions. The energy difference of the asymptotic values reflects the difference
of the ionization potentials of the two atoms. This feature is known as a \textit{step}
of the exact exchange-correlation potential in heteroatomic molecules
\cite{doi:10.1063/1.3271392}.
As seen from Fig.~\ref{fig:vxc_hetero_molec}~(b), the exchange-correlation potentials
of the $1/4$-interacting Kohn-Sham system (green-dashed line)
and that of the long-range interacting Kohn-Sham system (blue-dotted line) show
the weaker step feature than that of the non-interacting system (red-solid line).
To quantify the size of the step feature, we evaluate the difference of
the potential at $x=-8$~a.u. and $8$~a.u.,
$\Delta_{s}=v_{xc}(x=-8 \textrm{a.u.})-v_{xc}(x=8 \textrm{a.u.})$.
The step size of the non-interacting Kohn-Sham system is $\Delta_s = 0.47$~a.u.,
and that of the $1/4$-interacting Kohn-Sham system is $\Delta_s = 0.39$~a.u.
The long-range interacting Kohn-Sham system provides the smallest step size,
$\Delta_s = 0.15$~a.u.
Thus, one can clearly conclude that the effectively-interacting Kohn-Sham systems
significantly reduce the step feature of the exchange-correlation potential by
transferring a part of electronic correlation from the exchange-correlation potential
to the explicit correlation in the reference wavefunction.
Consistently with the above findings, the long-range interacting Kohn-Sham system
has the largest reduction of the complex feature of the exchange-correlation potential.
Thus, effective interaction based on the range separation is suggested to be a key
to achieve an efficient description of static and dynamical correlation with
the combined theory of DFT and wavefunction theory.

\section{Summary and Outlook\label{sec:summary}}

In order to explore a possibility to combine DFT and wavefunction
theory, we investigated a mapping from a fully interacting
problem to an effectively-interacting problem instead of
the conventional mapping to the non-interacting Kohn-Sham system.
To elucidate such mapping, we considered three kinds of
effectively-interacting Kohn-Sham systems.
One is the usual non-interacting Kohn-Sham system.
The second one is the $1/4$-interacting
Kohn-Sham system, where the effective interaction is set to the quarter
of the full interaction. The last one is the long-range interacting
Kohn-Sham system, where the short-range part of the full interaction
is ignored.

To practically investigate the properties of the effectively interacting
Kohn-Sham system, we first investigated the exact
exchange-correlation potentials of the one-dimensional
helium atom. As a result, we found that
the asymptotic behavior of the exchange-correlation potential
is determined by that of the residual interaction,
which is defined as the difference of the full and the effective
interactions.
This fact further indicates a possibility to construct a good
local density approximation for an effectively-interacting Kohn-Sham systems
by optimally choosing a short-range residual interaction
because the exchange-correlation potential vanishes as density vanishes
and the nonlocal density dependence is expected be suppressed.

Next, we evaluated the occupation distribution of the natural orbitals of
the ground-state wavefunction of the effectively-interacting Kohn-Sham systems.
As a result, we found that the occupations of the higher natural orbitals
are significantly suppressed in the effectively-interacting Kohn-Sham systems,
especially in the long-range interacting Kohn-Sham system.
This fact indicates that the effectively-interacting Kohn-Sham systems
offer an efficient decomposition of the electronic correlation
into the dynamical correlation in the DFT part and the static correlation
in the wavefunction part.

Then, we investigated the exact exchange-correlation potentials
of the one-dimensional hydrogen molecule.
Consistently with the previous works \cite{doi:10.1063/1.3271392,PhysRevB.93.155146,PhysRevA.54.1957,PhysRevB.90.241107},
we observed the spiky feature of the exchange-correlation potential. 
Once the effective interaction is turned on, the spiky feature is
strongly suppressed. Furthermore, in the long-range interacting Kohn-Sham system,
the spiky feature completely vanishes and the exchange-correlation potential
becomes smooth at the center of the molecule.

Finally, we investigated the exact exchange-correlation potentials of
the one-dimensional heteronuclear diatomic molecule in order to study
the step feature of the exchange-correlation potential,
which reflects the different ionization potentials
of the two atoms \cite{doi:10.1063/1.3271392}.
Consistently with the above analysis, we found that the difficult step feature
is significantly reduced in the effectively-interacting Kohn-Sham systems,
compared with the non-interacting Kohn-Sham system.
Especially, the long-range interacting Kohn-Sham system shows the smallest step feature,
indicating the effectiveness of the concept of the range separation
in the effectively-interacting Kohn-Sham systems.

Based on the above findings, we can conclude that 
the effectively-interacting Kohn-Sham approach
can open a way to efficiently describe the electronic correlation effect
by the combination of DFT and wavefunction theory,
decomposing the electronic correlation effect into
the DFT part and the wavefunction theory part. 

Another important fact is that the interacting Kohn-Sham scheme can be reduced
to the hybrid functional, applying the Hartree-Fock approximation to
the interacting Kohn-Sham equation, Eq.~(\ref{eq:ks-eq}).
Therefore, the interacting Kohn-Sham scheme offers a possibility 
to improve the hybrid functional approximation within the formal theoretical framework
of DFT, by adding the explicit correlation in the many-body Kohn-Sham wavefunctions.

In this work, we limited ourselves to the theoretical analysis of one-dimensional
two-electron systems in order to investigate the exact properties of
the interacting Kohn-Sham systems.
However, importantly, the above findings can be straightforwardly extended to
multi-dimensional many-electron systems 
and open a possibility to construct robust and efficient density functionals.
For example, based on the weak nonlocality of the exchange-correlation
potential of the interacting Kohn-Sham system, accurate LDA may be
constructed from the homogeneous electron gas with a suitable effective interaction.
Furthermore, inspired by the \textit{GW} method and the screened hybrid-functional approach,
a screening effect can be incorporated in an effective interaction of the Kohn-Sham system,
and accurate description for solid-state materials may be realized.
These extensions of the present work to multi-dimensional many-electron systems are
under way.

\begin{acknowledgments}
This work was supported by the European Research Council (ERC-2015-AdG694097),
the Cluster of Excellence 'Advanced Imaging of Matter' (AIM),
and JST-CREST under Grant No. JP-MJCR16N5.
Support by the Flatiron Institute, a division of the Simons Foundation is acknowledged.
S.A.S. gratefully acknowledges the fellowship from the Alexander von Humboldt Foundation.
\end{acknowledgments}

\bibliographystyle{apsrev4-1}
\bibliography{ref}

\begin{thebibliography}{48}%
\makeatletter
\providecommand \@ifxundefined [1]{%
 \@ifx{#1\undefined}
}%
\providecommand \@ifnum [1]{%
 \ifnum #1\expandafter \@firstoftwo
 \else \expandafter \@secondoftwo
 \fi
}%
\providecommand \@ifx [1]{%
 \ifx #1\expandafter \@firstoftwo
 \else \expandafter \@secondoftwo
 \fi
}%
\providecommand \natexlab [1]{#1}%
\providecommand \enquote  [1]{``#1''}%
\providecommand \bibnamefont  [1]{#1}%
\providecommand \bibfnamefont [1]{#1}%
\providecommand \citenamefont [1]{#1}%
\providecommand \href@noop [0]{\@secondoftwo}%
\providecommand \href [0]{\begingroup \@sanitize@url \@href}%
\providecommand \@href[1]{\@@startlink{#1}\@@href}%
\providecommand \@@href[1]{\endgroup#1\@@endlink}%
\providecommand \@sanitize@url [0]{\catcode `\\12\catcode `\$12\catcode
  `\&12\catcode `\#12\catcode `\^12\catcode `\_12\catcode `\%12\relax}%
\providecommand \@@startlink[1]{}%
\providecommand \@@endlink[0]{}%
\providecommand \url  [0]{\begingroup\@sanitize@url \@url }%
\providecommand \@url [1]{\endgroup\@href {#1}{\urlprefix }}%
\providecommand \urlprefix  [0]{URL }%
\providecommand \Eprint [0]{\href }%
\providecommand \doibase [0]{http://dx.doi.org/}%
\providecommand \selectlanguage [0]{\@gobble}%
\providecommand \bibinfo  [0]{\@secondoftwo}%
\providecommand \bibfield  [0]{\@secondoftwo}%
\providecommand \translation [1]{[#1]}%
\providecommand \BibitemOpen [0]{}%
\providecommand \bibitemStop [0]{}%
\providecommand \bibitemNoStop [0]{.\EOS\space}%
\providecommand \EOS [0]{\spacefactor3000\relax}%
\providecommand \BibitemShut  [1]{\csname bibitem#1\endcsname}%
\let\auto@bib@innerbib\@empty
\bibitem [{\citenamefont {Hohenberg}\ and\ \citenamefont
  {Kohn}(1964)}]{PhysRev.136.B864}%
  \BibitemOpen
  \bibfield  {author} {\bibinfo {author} {\bibfnamefont {P.}~\bibnamefont
  {Hohenberg}}\ and\ \bibinfo {author} {\bibfnamefont {W.}~\bibnamefont
  {Kohn}},\ }\href {\doibase 10.1103/PhysRev.136.B864} {\bibfield  {journal}
  {\bibinfo  {journal} {Phys. Rev.}\ }\textbf {\bibinfo {volume} {136}},\
  \bibinfo {pages} {B864} (\bibinfo {year} {1964})}\BibitemShut {NoStop}%
\bibitem [{\citenamefont {Kohn}\ and\ \citenamefont
  {Sham}(1965)}]{PhysRev.140.A1133}%
  \BibitemOpen
  \bibfield  {author} {\bibinfo {author} {\bibfnamefont {W.}~\bibnamefont
  {Kohn}}\ and\ \bibinfo {author} {\bibfnamefont {L.~J.}\ \bibnamefont
  {Sham}},\ }\href {\doibase 10.1103/PhysRev.140.A1133} {\bibfield  {journal}
  {\bibinfo  {journal} {Phys. Rev.}\ }\textbf {\bibinfo {volume} {140}},\
  \bibinfo {pages} {A1133} (\bibinfo {year} {1965})}\BibitemShut {NoStop}%
\bibitem [{\citenamefont {Perdew}\ and\ \citenamefont
  {Zunger}(1981)}]{PhysRevB.23.5048}%
  \BibitemOpen
  \bibfield  {author} {\bibinfo {author} {\bibfnamefont {J.~P.}\ \bibnamefont
  {Perdew}}\ and\ \bibinfo {author} {\bibfnamefont {A.}~\bibnamefont
  {Zunger}},\ }\href {\doibase 10.1103/PhysRevB.23.5048} {\bibfield  {journal}
  {\bibinfo  {journal} {Phys. Rev. B}\ }\textbf {\bibinfo {volume} {23}},\
  \bibinfo {pages} {5048} (\bibinfo {year} {1981})}\BibitemShut {NoStop}%
\bibitem [{\citenamefont {Perdew}\ and\ \citenamefont
  {Wang}(1992)}]{PhysRevB.45.13244}%
  \BibitemOpen
  \bibfield  {author} {\bibinfo {author} {\bibfnamefont {J.~P.}\ \bibnamefont
  {Perdew}}\ and\ \bibinfo {author} {\bibfnamefont {Y.}~\bibnamefont {Wang}},\
  }\href {\doibase 10.1103/PhysRevB.45.13244} {\bibfield  {journal} {\bibinfo
  {journal} {Phys. Rev. B}\ }\textbf {\bibinfo {volume} {45}},\ \bibinfo
  {pages} {13244} (\bibinfo {year} {1992})}\BibitemShut {NoStop}%
\bibitem [{\citenamefont {van Leeuwen}\ and\ \citenamefont
  {Baerends}(1994)}]{PhysRevA.49.2421}%
  \BibitemOpen
  \bibfield  {author} {\bibinfo {author} {\bibfnamefont {R.}~\bibnamefont {van
  Leeuwen}}\ and\ \bibinfo {author} {\bibfnamefont {E.~J.}\ \bibnamefont
  {Baerends}},\ }\href {\doibase 10.1103/PhysRevA.49.2421} {\bibfield
  {journal} {\bibinfo  {journal} {Phys. Rev. A}\ }\textbf {\bibinfo {volume}
  {49}},\ \bibinfo {pages} {2421} (\bibinfo {year} {1994})}\BibitemShut
  {NoStop}%
\bibitem [{\citenamefont {Perdew}\ \emph
  {et~al.}(1996{\natexlab{a}})\citenamefont {Perdew}, \citenamefont {Burke},\
  and\ \citenamefont {Ernzerhof}}]{PhysRevLett.77.3865}%
  \BibitemOpen
  \bibfield  {author} {\bibinfo {author} {\bibfnamefont {J.~P.}\ \bibnamefont
  {Perdew}}, \bibinfo {author} {\bibfnamefont {K.}~\bibnamefont {Burke}}, \
  and\ \bibinfo {author} {\bibfnamefont {M.}~\bibnamefont {Ernzerhof}},\ }\href
  {\doibase 10.1103/PhysRevLett.77.3865} {\bibfield  {journal} {\bibinfo
  {journal} {Phys. Rev. Lett.}\ }\textbf {\bibinfo {volume} {77}},\ \bibinfo
  {pages} {3865} (\bibinfo {year} {1996}{\natexlab{a}})}\BibitemShut {NoStop}%
\bibitem [{\citenamefont {Tao}\ \emph {et~al.}(2003)\citenamefont {Tao},
  \citenamefont {Perdew}, \citenamefont {Staroverov},\ and\ \citenamefont
  {Scuseria}}]{PhysRevLett.91.146401}%
  \BibitemOpen
  \bibfield  {author} {\bibinfo {author} {\bibfnamefont {J.}~\bibnamefont
  {Tao}}, \bibinfo {author} {\bibfnamefont {J.~P.}\ \bibnamefont {Perdew}},
  \bibinfo {author} {\bibfnamefont {V.~N.}\ \bibnamefont {Staroverov}}, \ and\
  \bibinfo {author} {\bibfnamefont {G.~E.}\ \bibnamefont {Scuseria}},\ }\href
  {\doibase 10.1103/PhysRevLett.91.146401} {\bibfield  {journal} {\bibinfo
  {journal} {Phys. Rev. Lett.}\ }\textbf {\bibinfo {volume} {91}},\ \bibinfo
  {pages} {146401} (\bibinfo {year} {2003})}\BibitemShut {NoStop}%
\bibitem [{\citenamefont {Van~Voorhis}\ and\ \citenamefont
  {Scuseria}(1998)}]{doi:10.1063/1.476577}%
  \BibitemOpen
  \bibfield  {author} {\bibinfo {author} {\bibfnamefont {T.}~\bibnamefont
  {Van~Voorhis}}\ and\ \bibinfo {author} {\bibfnamefont {G.~E.}\ \bibnamefont
  {Scuseria}},\ }\href {\doibase 10.1063/1.476577} {\bibfield  {journal}
  {\bibinfo  {journal} {The Journal of Chemical Physics}\ }\textbf {\bibinfo
  {volume} {109}},\ \bibinfo {pages} {400} (\bibinfo {year}
  {1998})}\BibitemShut {NoStop}%
\bibitem [{\citenamefont {Sun}\ \emph {et~al.}(2015)\citenamefont {Sun},
  \citenamefont {Ruzsinszky},\ and\ \citenamefont
  {Perdew}}]{PhysRevLett.115.036402}%
  \BibitemOpen
  \bibfield  {author} {\bibinfo {author} {\bibfnamefont {J.}~\bibnamefont
  {Sun}}, \bibinfo {author} {\bibfnamefont {A.}~\bibnamefont {Ruzsinszky}}, \
  and\ \bibinfo {author} {\bibfnamefont {J.~P.}\ \bibnamefont {Perdew}},\
  }\href {\doibase 10.1103/PhysRevLett.115.036402} {\bibfield  {journal}
  {\bibinfo  {journal} {Phys. Rev. Lett.}\ }\textbf {\bibinfo {volume} {115}},\
  \bibinfo {pages} {036402} (\bibinfo {year} {2015})}\BibitemShut {NoStop}%
\bibitem [{\citenamefont {Becke}(1993)}]{doi:10.1063/1.464913}%
  \BibitemOpen
  \bibfield  {author} {\bibinfo {author} {\bibfnamefont {A.~D.}\ \bibnamefont
  {Becke}},\ }\href {\doibase 10.1063/1.464913} {\bibfield  {journal} {\bibinfo
   {journal} {The Journal of Chemical Physics}\ }\textbf {\bibinfo {volume}
  {98}},\ \bibinfo {pages} {5648} (\bibinfo {year} {1993})}\BibitemShut
  {NoStop}%
\bibitem [{\citenamefont {Perdew}\ \emph
  {et~al.}(1996{\natexlab{b}})\citenamefont {Perdew}, \citenamefont
  {Ernzerhof},\ and\ \citenamefont {Burke}}]{doi:10.1063/1.472933}%
  \BibitemOpen
  \bibfield  {author} {\bibinfo {author} {\bibfnamefont {J.~P.}\ \bibnamefont
  {Perdew}}, \bibinfo {author} {\bibfnamefont {M.}~\bibnamefont {Ernzerhof}}, \
  and\ \bibinfo {author} {\bibfnamefont {K.}~\bibnamefont {Burke}},\ }\href
  {\doibase 10.1063/1.472933} {\bibfield  {journal} {\bibinfo  {journal} {The
  Journal of Chemical Physics}\ }\textbf {\bibinfo {volume} {105}},\ \bibinfo
  {pages} {9982} (\bibinfo {year} {1996}{\natexlab{b}})}\BibitemShut {NoStop}%
\bibitem [{\citenamefont {Heyd}\ \emph {et~al.}(2003)\citenamefont {Heyd},
  \citenamefont {Scuseria},\ and\ \citenamefont
  {Ernzerhof}}]{doi:10.1063/1.1564060}%
  \BibitemOpen
  \bibfield  {author} {\bibinfo {author} {\bibfnamefont {J.}~\bibnamefont
  {Heyd}}, \bibinfo {author} {\bibfnamefont {G.~E.}\ \bibnamefont {Scuseria}},
  \ and\ \bibinfo {author} {\bibfnamefont {M.}~\bibnamefont {Ernzerhof}},\
  }\href {\doibase 10.1063/1.1564060} {\bibfield  {journal} {\bibinfo
  {journal} {The Journal of Chemical Physics}\ }\textbf {\bibinfo {volume}
  {118}},\ \bibinfo {pages} {8207} (\bibinfo {year} {2003})}\BibitemShut
  {NoStop}%
\bibitem [{\citenamefont {Levy}\ \emph {et~al.}(1984)\citenamefont {Levy},
  \citenamefont {Perdew},\ and\ \citenamefont {Sahni}}]{PhysRevA.30.2745}%
  \BibitemOpen
  \bibfield  {author} {\bibinfo {author} {\bibfnamefont {M.}~\bibnamefont
  {Levy}}, \bibinfo {author} {\bibfnamefont {J.~P.}\ \bibnamefont {Perdew}}, \
  and\ \bibinfo {author} {\bibfnamefont {V.}~\bibnamefont {Sahni}},\ }\href
  {\doibase 10.1103/PhysRevA.30.2745} {\bibfield  {journal} {\bibinfo
  {journal} {Phys. Rev. A}\ }\textbf {\bibinfo {volume} {30}},\ \bibinfo
  {pages} {2745} (\bibinfo {year} {1984})}\BibitemShut {NoStop}%
\bibitem [{\citenamefont {Almbladh}\ and\ \citenamefont
  {Pedroza}(1984)}]{PhysRevA.29.2322}%
  \BibitemOpen
  \bibfield  {author} {\bibinfo {author} {\bibfnamefont {C.~O.}\ \bibnamefont
  {Almbladh}}\ and\ \bibinfo {author} {\bibfnamefont {A.~C.}\ \bibnamefont
  {Pedroza}},\ }\href {\doibase 10.1103/PhysRevA.29.2322} {\bibfield  {journal}
  {\bibinfo  {journal} {Phys. Rev. A}\ }\textbf {\bibinfo {volume} {29}},\
  \bibinfo {pages} {2322} (\bibinfo {year} {1984})}\BibitemShut {NoStop}%
\bibitem [{\citenamefont {Buijse}\ \emph {et~al.}(1989)\citenamefont {Buijse},
  \citenamefont {Baerends},\ and\ \citenamefont {Snijders}}]{PhysRevA.40.4190}%
  \BibitemOpen
  \bibfield  {author} {\bibinfo {author} {\bibfnamefont {M.~A.}\ \bibnamefont
  {Buijse}}, \bibinfo {author} {\bibfnamefont {E.~J.}\ \bibnamefont
  {Baerends}}, \ and\ \bibinfo {author} {\bibfnamefont {J.~G.}\ \bibnamefont
  {Snijders}},\ }\href {\doibase 10.1103/PhysRevA.40.4190} {\bibfield
  {journal} {\bibinfo  {journal} {Phys. Rev. A}\ }\textbf {\bibinfo {volume}
  {40}},\ \bibinfo {pages} {4190} (\bibinfo {year} {1989})}\BibitemShut
  {NoStop}%
\bibitem [{\citenamefont {Gritsenko}\ and\ \citenamefont
  {Baerends}(1997)}]{Gritsenko1997}%
  \BibitemOpen
  \bibfield  {author} {\bibinfo {author} {\bibfnamefont {O.~V.}\ \bibnamefont
  {Gritsenko}}\ and\ \bibinfo {author} {\bibfnamefont {E.~J.}\ \bibnamefont
  {Baerends}},\ }\href {\doibase 10.1007/s002140050202} {\bibfield  {journal}
  {\bibinfo  {journal} {Theoretical Chemistry Accounts}\ }\textbf {\bibinfo
  {volume} {96}},\ \bibinfo {pages} {44} (\bibinfo {year} {1997})}\BibitemShut
  {NoStop}%
\bibitem [{\citenamefont {Helbig}\ \emph {et~al.}(2009)\citenamefont {Helbig},
  \citenamefont {Tokatly},\ and\ \citenamefont
  {Rubio}}]{doi:10.1063/1.3271392}%
  \BibitemOpen
  \bibfield  {author} {\bibinfo {author} {\bibfnamefont {N.}~\bibnamefont
  {Helbig}}, \bibinfo {author} {\bibfnamefont {I.~V.}\ \bibnamefont {Tokatly}},
  \ and\ \bibinfo {author} {\bibfnamefont {A.}~\bibnamefont {Rubio}},\ }\href
  {\doibase 10.1063/1.3271392} {\bibfield  {journal} {\bibinfo  {journal} {The
  Journal of Chemical Physics}\ }\textbf {\bibinfo {volume} {131}},\ \bibinfo
  {pages} {224105} (\bibinfo {year} {2009})}\BibitemShut {NoStop}%
\bibitem [{\citenamefont {Hodgson}\ \emph {et~al.}(2016)\citenamefont
  {Hodgson}, \citenamefont {Ramsden},\ and\ \citenamefont
  {Godby}}]{PhysRevB.93.155146}%
  \BibitemOpen
  \bibfield  {author} {\bibinfo {author} {\bibfnamefont {M.~J.~P.}\
  \bibnamefont {Hodgson}}, \bibinfo {author} {\bibfnamefont {J.~D.}\
  \bibnamefont {Ramsden}}, \ and\ \bibinfo {author} {\bibfnamefont {R.~W.}\
  \bibnamefont {Godby}},\ }\href {\doibase 10.1103/PhysRevB.93.155146}
  {\bibfield  {journal} {\bibinfo  {journal} {Phys. Rev. B}\ }\textbf {\bibinfo
  {volume} {93}},\ \bibinfo {pages} {155146} (\bibinfo {year}
  {2016})}\BibitemShut {NoStop}%
\bibitem [{\citenamefont {Burke}(2006)}]{Burke2006exactconditions}%
  \BibitemOpen
  \bibfield  {author} {\bibinfo {author} {\bibfnamefont {K.}~\bibnamefont
  {Burke}},\ }\href {\doibase 10.1007/3-540-35426-3 11} {\bibfield  {journal}
  {\bibinfo  {journal} {Lect. Notes Phys.}\ }\textbf {\bibinfo {volume}
  {706}},\ \bibinfo {pages} {181} (\bibinfo {year} {2006})}\BibitemShut
  {NoStop}%
\bibitem [{\citenamefont {Perdew}\ and\ \citenamefont
  {Schmidt}(2001)}]{doi:10.1063/1.1390175}%
  \BibitemOpen
  \bibfield  {author} {\bibinfo {author} {\bibfnamefont {J.~P.}\ \bibnamefont
  {Perdew}}\ and\ \bibinfo {author} {\bibfnamefont {K.}~\bibnamefont
  {Schmidt}},\ }\href {\doibase 10.1063/1.1390175} {\bibfield  {journal}
  {\bibinfo  {journal} {AIP Conference Proceedings}\ }\textbf {\bibinfo
  {volume} {577}},\ \bibinfo {pages} {1} (\bibinfo {year} {2001})}\BibitemShut
  {NoStop}%
\bibitem [{\citenamefont {Medvedev}\ \emph {et~al.}(2017)\citenamefont
  {Medvedev}, \citenamefont {Bushmarinov}, \citenamefont {Sun}, \citenamefont
  {Perdew},\ and\ \citenamefont {Lyssenko}}]{Medvedev49}%
  \BibitemOpen
  \bibfield  {author} {\bibinfo {author} {\bibfnamefont {M.~G.}\ \bibnamefont
  {Medvedev}}, \bibinfo {author} {\bibfnamefont {I.~S.}\ \bibnamefont
  {Bushmarinov}}, \bibinfo {author} {\bibfnamefont {J.}~\bibnamefont {Sun}},
  \bibinfo {author} {\bibfnamefont {J.~P.}\ \bibnamefont {Perdew}}, \ and\
  \bibinfo {author} {\bibfnamefont {K.~A.}\ \bibnamefont {Lyssenko}},\ }\href
  {\doibase 10.1126/science.aah5975} {\bibfield  {journal} {\bibinfo  {journal}
  {Science}\ }\textbf {\bibinfo {volume} {355}},\ \bibinfo {pages} {49}
  (\bibinfo {year} {2017})}\BibitemShut {NoStop}%
\bibitem [{\citenamefont {Szabo}\ and\ \citenamefont
  {Ostlund}(1989)}]{szabo1989book}%
  \BibitemOpen
  \bibfield  {author} {\bibinfo {author} {\bibfnamefont {A.}~\bibnamefont
  {Szabo}}\ and\ \bibinfo {author} {\bibfnamefont {N.~S.}\ \bibnamefont
  {Ostlund}},\ }\href@noop {} {\emph {\bibinfo {title} {Modern Quantum
  Chemistry}}}\ (\bibinfo  {publisher} {McGraw-Hill, New York},\ \bibinfo
  {year} {1989})\BibitemShut {NoStop}%
\bibitem [{\citenamefont {Aryasetiawan}\ and\ \citenamefont
  {Gunnarsson}(1998)}]{Aryasetiawan_1998}%
  \BibitemOpen
  \bibfield  {author} {\bibinfo {author} {\bibfnamefont {F.}~\bibnamefont
  {Aryasetiawan}}\ and\ \bibinfo {author} {\bibfnamefont {O.}~\bibnamefont
  {Gunnarsson}},\ }\href {\doibase 10.1088/0034-4885/61/3/002} {\bibfield
  {journal} {\bibinfo  {journal} {Reports on Progress in Physics}\ }\textbf
  {\bibinfo {volume} {61}},\ \bibinfo {pages} {237} (\bibinfo {year}
  {1998})}\BibitemShut {NoStop}%
\bibitem [{\citenamefont {Aulbur}\ \emph {et~al.}(2000)\citenamefont {Aulbur},
  \citenamefont {Jönsson},\ and\ \citenamefont {Wilkins}}]{AULBUR20001}%
  \BibitemOpen
  \bibfield  {author} {\bibinfo {author} {\bibfnamefont {W.~G.}\ \bibnamefont
  {Aulbur}}, \bibinfo {author} {\bibfnamefont {L.}~\bibnamefont {Jönsson}}, \
  and\ \bibinfo {author} {\bibfnamefont {J.~W.}\ \bibnamefont {Wilkins}}\
  }(\bibinfo  {publisher} {Academic Press},\ \bibinfo {year} {2000})\ pp.\
  \bibinfo {pages} {1 -- 218}\BibitemShut {NoStop}%
\bibitem [{\citenamefont {Onida}\ \emph {et~al.}(2002)\citenamefont {Onida},
  \citenamefont {Reining},\ and\ \citenamefont {Rubio}}]{RevModPhys.74.601}%
  \BibitemOpen
  \bibfield  {author} {\bibinfo {author} {\bibfnamefont {G.}~\bibnamefont
  {Onida}}, \bibinfo {author} {\bibfnamefont {L.}~\bibnamefont {Reining}}, \
  and\ \bibinfo {author} {\bibfnamefont {A.}~\bibnamefont {Rubio}},\ }\href
  {\doibase 10.1103/RevModPhys.74.601} {\bibfield  {journal} {\bibinfo
  {journal} {Rev. Mod. Phys.}\ }\textbf {\bibinfo {volume} {74}},\ \bibinfo
  {pages} {601} (\bibinfo {year} {2002})}\BibitemShut {NoStop}%
\bibitem [{\citenamefont {van Setten}\ \emph {et~al.}(2013)\citenamefont {van
  Setten}, \citenamefont {Weigend},\ and\ \citenamefont
  {Evers}}]{vanSetten2013}%
  \BibitemOpen
  \bibfield  {author} {\bibinfo {author} {\bibfnamefont {M.~J.}\ \bibnamefont
  {van Setten}}, \bibinfo {author} {\bibfnamefont {F.}~\bibnamefont {Weigend}},
  \ and\ \bibinfo {author} {\bibfnamefont {F.}~\bibnamefont {Evers}},\ }\href
  {\doibase 10.1021/ct300648t} {\bibfield  {journal} {\bibinfo  {journal}
  {Journal of Chemical Theory and Computation}\ }\textbf {\bibinfo {volume}
  {9}},\ \bibinfo {pages} {232} (\bibinfo {year} {2013})}\BibitemShut {NoStop}%
\bibitem [{\citenamefont {Foulkes}\ \emph {et~al.}(2001)\citenamefont
  {Foulkes}, \citenamefont {Mitas}, \citenamefont {Needs},\ and\ \citenamefont
  {Rajagopal}}]{RevModPhys.73.33}%
  \BibitemOpen
  \bibfield  {author} {\bibinfo {author} {\bibfnamefont {W.~M.~C.}\
  \bibnamefont {Foulkes}}, \bibinfo {author} {\bibfnamefont {L.}~\bibnamefont
  {Mitas}}, \bibinfo {author} {\bibfnamefont {R.~J.}\ \bibnamefont {Needs}}, \
  and\ \bibinfo {author} {\bibfnamefont {G.}~\bibnamefont {Rajagopal}},\ }\href
  {\doibase 10.1103/RevModPhys.73.33} {\bibfield  {journal} {\bibinfo
  {journal} {Rev. Mod. Phys.}\ }\textbf {\bibinfo {volume} {73}},\ \bibinfo
  {pages} {33} (\bibinfo {year} {2001})}\BibitemShut {NoStop}%
\bibitem [{\citenamefont {Stella}\ \emph {et~al.}(2011)\citenamefont {Stella},
  \citenamefont {Attaccalite}, \citenamefont {Sorella},\ and\ \citenamefont
  {Rubio}}]{PhysRevB.84.245117}%
  \BibitemOpen
  \bibfield  {author} {\bibinfo {author} {\bibfnamefont {L.}~\bibnamefont
  {Stella}}, \bibinfo {author} {\bibfnamefont {C.}~\bibnamefont {Attaccalite}},
  \bibinfo {author} {\bibfnamefont {S.}~\bibnamefont {Sorella}}, \ and\
  \bibinfo {author} {\bibfnamefont {A.}~\bibnamefont {Rubio}},\ }\href
  {\doibase 10.1103/PhysRevB.84.245117} {\bibfield  {journal} {\bibinfo
  {journal} {Phys. Rev. B}\ }\textbf {\bibinfo {volume} {84}},\ \bibinfo
  {pages} {245117} (\bibinfo {year} {2011})}\BibitemShut {NoStop}%
\bibitem [{\citenamefont {Hunt}\ \emph {et~al.}(2018)\citenamefont {Hunt},
  \citenamefont {Szyniszewski}, \citenamefont {Prayogo}, \citenamefont
  {Maezono},\ and\ \citenamefont {Drummond}}]{PhysRevB.98.075122}%
  \BibitemOpen
  \bibfield  {author} {\bibinfo {author} {\bibfnamefont {R.~J.}\ \bibnamefont
  {Hunt}}, \bibinfo {author} {\bibfnamefont {M.}~\bibnamefont {Szyniszewski}},
  \bibinfo {author} {\bibfnamefont {G.~I.}\ \bibnamefont {Prayogo}}, \bibinfo
  {author} {\bibfnamefont {R.}~\bibnamefont {Maezono}}, \ and\ \bibinfo
  {author} {\bibfnamefont {N.~D.}\ \bibnamefont {Drummond}},\ }\href {\doibase
  10.1103/PhysRevB.98.075122} {\bibfield  {journal} {\bibinfo  {journal} {Phys.
  Rev. B}\ }\textbf {\bibinfo {volume} {98}},\ \bibinfo {pages} {075122}
  (\bibinfo {year} {2018})}\BibitemShut {NoStop}%
\bibitem [{\citenamefont {Mazzola}\ \emph {et~al.}(2018)\citenamefont
  {Mazzola}, \citenamefont {Helled},\ and\ \citenamefont
  {Sorella}}]{PhysRevLett.120.025701}%
  \BibitemOpen
  \bibfield  {author} {\bibinfo {author} {\bibfnamefont {G.}~\bibnamefont
  {Mazzola}}, \bibinfo {author} {\bibfnamefont {R.}~\bibnamefont {Helled}}, \
  and\ \bibinfo {author} {\bibfnamefont {S.}~\bibnamefont {Sorella}},\ }\href
  {\doibase 10.1103/PhysRevLett.120.025701} {\bibfield  {journal} {\bibinfo
  {journal} {Phys. Rev. Lett.}\ }\textbf {\bibinfo {volume} {120}},\ \bibinfo
  {pages} {025701} (\bibinfo {year} {2018})}\BibitemShut {NoStop}%
\bibitem [{\citenamefont {Gritsenko}\ \emph {et~al.}(1997)\citenamefont
  {Gritsenko}, \citenamefont {Schipper},\ and\ \citenamefont
  {Baerends}}]{doi:10.1063/1.474864}%
  \BibitemOpen
  \bibfield  {author} {\bibinfo {author} {\bibfnamefont {O.~V.}\ \bibnamefont
  {Gritsenko}}, \bibinfo {author} {\bibfnamefont {P.~R.~T.}\ \bibnamefont
  {Schipper}}, \ and\ \bibinfo {author} {\bibfnamefont {E.~J.}\ \bibnamefont
  {Baerends}},\ }\href {\doibase 10.1063/1.474864} {\bibfield  {journal}
  {\bibinfo  {journal} {The Journal of Chemical Physics}\ }\textbf {\bibinfo
  {volume} {107}},\ \bibinfo {pages} {5007} (\bibinfo {year}
  {1997})}\BibitemShut {NoStop}%
\bibitem [{\citenamefont {Cohen}\ \emph {et~al.}(2008)\citenamefont {Cohen},
  \citenamefont {Mori-S{\'a}nchez},\ and\ \citenamefont {Yang}}]{Cohen792}%
  \BibitemOpen
  \bibfield  {author} {\bibinfo {author} {\bibfnamefont {A.~J.}\ \bibnamefont
  {Cohen}}, \bibinfo {author} {\bibfnamefont {P.}~\bibnamefont
  {Mori-S{\'a}nchez}}, \ and\ \bibinfo {author} {\bibfnamefont
  {W.}~\bibnamefont {Yang}},\ }\href {\doibase 10.1126/science.1158722}
  {\bibfield  {journal} {\bibinfo  {journal} {Science}\ }\textbf {\bibinfo
  {volume} {321}},\ \bibinfo {pages} {792} (\bibinfo {year}
  {2008})}\BibitemShut {NoStop}%
\bibitem [{\citenamefont {Mori-S\'anchez}\ \emph {et~al.}(2009)\citenamefont
  {Mori-S\'anchez}, \citenamefont {Cohen},\ and\ \citenamefont
  {Yang}}]{PhysRevLett.102.066403}%
  \BibitemOpen
  \bibfield  {author} {\bibinfo {author} {\bibfnamefont {P.}~\bibnamefont
  {Mori-S\'anchez}}, \bibinfo {author} {\bibfnamefont {A.~J.}\ \bibnamefont
  {Cohen}}, \ and\ \bibinfo {author} {\bibfnamefont {W.}~\bibnamefont {Yang}},\
  }\href {\doibase 10.1103/PhysRevLett.102.066403} {\bibfield  {journal}
  {\bibinfo  {journal} {Phys. Rev. Lett.}\ }\textbf {\bibinfo {volume} {102}},\
  \bibinfo {pages} {066403} (\bibinfo {year} {2009})}\BibitemShut {NoStop}%
\bibitem [{\citenamefont {Ess}\ \emph {et~al.}(2011)\citenamefont {Ess},
  \citenamefont {Johnson}, \citenamefont {Hu},\ and\ \citenamefont
  {Yang}}]{Ess2011}%
  \BibitemOpen
  \bibfield  {author} {\bibinfo {author} {\bibfnamefont {D.~H.}\ \bibnamefont
  {Ess}}, \bibinfo {author} {\bibfnamefont {E.~R.}\ \bibnamefont {Johnson}},
  \bibinfo {author} {\bibfnamefont {X.}~\bibnamefont {Hu}}, \ and\ \bibinfo
  {author} {\bibfnamefont {W.}~\bibnamefont {Yang}},\ }\href {\doibase
  10.1021/jp109280y} {\bibfield  {journal} {\bibinfo  {journal} {The Journal of
  Physical Chemistry A}\ }\textbf {\bibinfo {volume} {115}},\ \bibinfo {pages}
  {76} (\bibinfo {year} {2011})}\BibitemShut {NoStop}%
\bibitem [{\citenamefont {Seidl}\ \emph {et~al.}(1996)\citenamefont {Seidl},
  \citenamefont {G\"orling}, \citenamefont {Vogl}, \citenamefont {Majewski},\
  and\ \citenamefont {Levy}}]{PhysRevB.53.3764}%
  \BibitemOpen
  \bibfield  {author} {\bibinfo {author} {\bibfnamefont {A.}~\bibnamefont
  {Seidl}}, \bibinfo {author} {\bibfnamefont {A.}~\bibnamefont {G\"orling}},
  \bibinfo {author} {\bibfnamefont {P.}~\bibnamefont {Vogl}}, \bibinfo {author}
  {\bibfnamefont {J.~A.}\ \bibnamefont {Majewski}}, \ and\ \bibinfo {author}
  {\bibfnamefont {M.}~\bibnamefont {Levy}},\ }\href {\doibase
  10.1103/PhysRevB.53.3764} {\bibfield  {journal} {\bibinfo  {journal} {Phys.
  Rev. B}\ }\textbf {\bibinfo {volume} {53}},\ \bibinfo {pages} {3764}
  (\bibinfo {year} {1996})}\BibitemShut {NoStop}%
\bibitem [{\citenamefont {Grimme}(1996)}]{GRIMME1996128}%
  \BibitemOpen
  \bibfield  {author} {\bibinfo {author} {\bibfnamefont {S.}~\bibnamefont
  {Grimme}},\ }\href {\doibase https://doi.org/10.1016/0009-2614(96)00722-1}
  {\bibfield  {journal} {\bibinfo  {journal} {Chemical Physics Letters}\
  }\textbf {\bibinfo {volume} {259}},\ \bibinfo {pages} {128 } (\bibinfo {year}
  {1996})}\BibitemShut {NoStop}%
\bibitem [{\citenamefont {Borowski}\ \emph {et~al.}(1998)\citenamefont
  {Borowski}, \citenamefont {Jordan}, \citenamefont {Nichols},\ and\
  \citenamefont {Nachtigall}}]{Borowski1998}%
  \BibitemOpen
  \bibfield  {author} {\bibinfo {author} {\bibfnamefont {P.}~\bibnamefont
  {Borowski}}, \bibinfo {author} {\bibfnamefont {K.~D.}\ \bibnamefont
  {Jordan}}, \bibinfo {author} {\bibfnamefont {J.}~\bibnamefont {Nichols}}, \
  and\ \bibinfo {author} {\bibfnamefont {P.}~\bibnamefont {Nachtigall}},\
  }\href {\doibase 10.1007/s002140050315} {\bibfield  {journal} {\bibinfo
  {journal} {Theoretical Chemistry Accounts}\ }\textbf {\bibinfo {volume}
  {99}},\ \bibinfo {pages} {135} (\bibinfo {year} {1998})}\BibitemShut
  {NoStop}%
\bibitem [{\citenamefont {Grimme}\ and\ \citenamefont
  {Waletzke}(1999)}]{doi:10.1063/1.479866}%
  \BibitemOpen
  \bibfield  {author} {\bibinfo {author} {\bibfnamefont {S.}~\bibnamefont
  {Grimme}}\ and\ \bibinfo {author} {\bibfnamefont {M.}~\bibnamefont
  {Waletzke}},\ }\href {\doibase 10.1063/1.479866} {\bibfield  {journal}
  {\bibinfo  {journal} {The Journal of Chemical Physics}\ }\textbf {\bibinfo
  {volume} {111}},\ \bibinfo {pages} {5645} (\bibinfo {year}
  {1999})}\BibitemShut {NoStop}%
\bibitem [{\citenamefont {Gr{\"a}fenstein}\ \emph {et~al.}(1998)\citenamefont
  {Gr{\"a}fenstein}, \citenamefont {Kraka},\ and\ \citenamefont
  {Cremer}}]{GRAFENSTEIN1998593}%
  \BibitemOpen
  \bibfield  {author} {\bibinfo {author} {\bibfnamefont {J.}~\bibnamefont
  {Gr{\"a}fenstein}}, \bibinfo {author} {\bibfnamefont {E.}~\bibnamefont
  {Kraka}}, \ and\ \bibinfo {author} {\bibfnamefont {D.}~\bibnamefont
  {Cremer}},\ }\href {\doibase https://doi.org/10.1016/S0009-2614(98)00335-2}
  {\bibfield  {journal} {\bibinfo  {journal} {Chemical Physics Letters}\
  }\textbf {\bibinfo {volume} {288}},\ \bibinfo {pages} {593 } (\bibinfo {year}
  {1998})}\BibitemShut {NoStop}%
\bibitem [{\citenamefont {Filatov}\ and\ \citenamefont
  {Shaik}(1998)}]{FILATOV1998689}%
  \BibitemOpen
  \bibfield  {author} {\bibinfo {author} {\bibfnamefont {M.}~\bibnamefont
  {Filatov}}\ and\ \bibinfo {author} {\bibfnamefont {S.}~\bibnamefont
  {Shaik}},\ }\href {\doibase https://doi.org/10.1016/S0009-2614(98)00364-9}
  {\bibfield  {journal} {\bibinfo  {journal} {Chemical Physics Letters}\
  }\textbf {\bibinfo {volume} {288}},\ \bibinfo {pages} {689 } (\bibinfo {year}
  {1998})}\BibitemShut {NoStop}%
\bibitem [{\citenamefont {Roemelt}\ \emph {et~al.}(2013)\citenamefont
  {Roemelt}, \citenamefont {Maganas}, \citenamefont {DeBeer},\ and\
  \citenamefont {Neese}}]{doi:10.1063/1.4804607}%
  \BibitemOpen
  \bibfield  {author} {\bibinfo {author} {\bibfnamefont {M.}~\bibnamefont
  {Roemelt}}, \bibinfo {author} {\bibfnamefont {D.}~\bibnamefont {Maganas}},
  \bibinfo {author} {\bibfnamefont {S.}~\bibnamefont {DeBeer}}, \ and\ \bibinfo
  {author} {\bibfnamefont {F.}~\bibnamefont {Neese}},\ }\href {\doibase
  10.1063/1.4804607} {\bibfield  {journal} {\bibinfo  {journal} {The Journal of
  Chemical Physics}\ }\textbf {\bibinfo {volume} {138}},\ \bibinfo {pages}
  {204101} (\bibinfo {year} {2013})}\BibitemShut {NoStop}%
\bibitem [{\citenamefont {Fromager}\ \emph {et~al.}(2007)\citenamefont
  {Fromager}, \citenamefont {Toulouse},\ and\ \citenamefont
  {Jensen}}]{doi:10.1063/1.2566459}%
  \BibitemOpen
  \bibfield  {author} {\bibinfo {author} {\bibfnamefont {E.}~\bibnamefont
  {Fromager}}, \bibinfo {author} {\bibfnamefont {J.}~\bibnamefont {Toulouse}},
  \ and\ \bibinfo {author} {\bibfnamefont {H.~J.~A.}\ \bibnamefont {Jensen}},\
  }\href {\doibase 10.1063/1.2566459} {\bibfield  {journal} {\bibinfo
  {journal} {The Journal of Chemical Physics}\ }\textbf {\bibinfo {volume}
  {126}},\ \bibinfo {pages} {074111} (\bibinfo {year} {2007})}\BibitemShut
  {NoStop}%
\bibitem [{\citenamefont {Levy}\ and\ \citenamefont
  {Zahariev}(2014)}]{PhysRevLett.113.113002}%
  \BibitemOpen
  \bibfield  {author} {\bibinfo {author} {\bibfnamefont {M.}~\bibnamefont
  {Levy}}\ and\ \bibinfo {author} {\bibfnamefont {F.}~\bibnamefont
  {Zahariev}},\ }\href {\doibase 10.1103/PhysRevLett.113.113002} {\bibfield
  {journal} {\bibinfo  {journal} {Phys. Rev. Lett.}\ }\textbf {\bibinfo
  {volume} {113}},\ \bibinfo {pages} {113002} (\bibinfo {year}
  {2014})}\BibitemShut {NoStop}%
\bibitem [{\citenamefont {Wang}\ and\ \citenamefont
  {Parr}(1993)}]{PhysRevA.47.R1591}%
  \BibitemOpen
  \bibfield  {author} {\bibinfo {author} {\bibfnamefont {Y.}~\bibnamefont
  {Wang}}\ and\ \bibinfo {author} {\bibfnamefont {R.~G.}\ \bibnamefont
  {Parr}},\ }\href {\doibase 10.1103/PhysRevA.47.R1591} {\bibfield  {journal}
  {\bibinfo  {journal} {Phys. Rev. A}\ }\textbf {\bibinfo {volume} {47}},\
  \bibinfo {pages} {R1591} (\bibinfo {year} {1993})}\BibitemShut {NoStop}%
\bibitem [{\citenamefont {Nielsen}\ \emph {et~al.}(2018)\citenamefont
  {Nielsen}, \citenamefont {Ruggenthaler},\ and\ \citenamefont {van
  Leeuwen}}]{Nielsen2018}%
  \BibitemOpen
  \bibfield  {author} {\bibinfo {author} {\bibfnamefont {S.~E.~B.}\
  \bibnamefont {Nielsen}}, \bibinfo {author} {\bibfnamefont {M.}~\bibnamefont
  {Ruggenthaler}}, \ and\ \bibinfo {author} {\bibfnamefont {R.}~\bibnamefont
  {van Leeuwen}},\ }\href {\doibase 10.1140/epjb/e2018-90276-4} {\bibfield
  {journal} {\bibinfo  {journal} {The European Physical Journal B}\ }\textbf
  {\bibinfo {volume} {91}},\ \bibinfo {pages} {235} (\bibinfo {year}
  {2018})}\BibitemShut {NoStop}%
\bibitem [{\citenamefont {L\"owdin}(1955)}]{PhysRev.97.1474}%
  \BibitemOpen
  \bibfield  {author} {\bibinfo {author} {\bibfnamefont {P.-O.}\ \bibnamefont
  {L\"owdin}},\ }\href {\doibase 10.1103/PhysRev.97.1474} {\bibfield  {journal}
  {\bibinfo  {journal} {Phys. Rev.}\ }\textbf {\bibinfo {volume} {97}},\
  \bibinfo {pages} {1474} (\bibinfo {year} {1955})}\BibitemShut {NoStop}%
\bibitem [{\citenamefont {Gritsenko}\ and\ \citenamefont
  {Baerends}(1996)}]{PhysRevA.54.1957}%
  \BibitemOpen
  \bibfield  {author} {\bibinfo {author} {\bibfnamefont {O.~V.}\ \bibnamefont
  {Gritsenko}}\ and\ \bibinfo {author} {\bibfnamefont {E.~J.}\ \bibnamefont
  {Baerends}},\ }\href {\doibase 10.1103/PhysRevA.54.1957} {\bibfield
  {journal} {\bibinfo  {journal} {Phys. Rev. A}\ }\textbf {\bibinfo {volume}
  {54}},\ \bibinfo {pages} {1957} (\bibinfo {year} {1996})}\BibitemShut
  {NoStop}%
\bibitem [{\citenamefont {Hodgson}\ \emph {et~al.}(2014)\citenamefont
  {Hodgson}, \citenamefont {Ramsden}, \citenamefont {Durrant},\ and\
  \citenamefont {Godby}}]{PhysRevB.90.241107}%
  \BibitemOpen
  \bibfield  {author} {\bibinfo {author} {\bibfnamefont {M.~J.~P.}\
  \bibnamefont {Hodgson}}, \bibinfo {author} {\bibfnamefont {J.~D.}\
  \bibnamefont {Ramsden}}, \bibinfo {author} {\bibfnamefont {T.~R.}\
  \bibnamefont {Durrant}}, \ and\ \bibinfo {author} {\bibfnamefont {R.~W.}\
  \bibnamefont {Godby}},\ }\href {\doibase 10.1103/PhysRevB.90.241107}
  {\bibfield  {journal} {\bibinfo  {journal} {Phys. Rev. B}\ }\textbf {\bibinfo
  {volume} {90}},\ \bibinfo {pages} {241107} (\bibinfo {year}
  {2014})}\BibitemShut {NoStop}%
\end{thebibliography}%

\end{document}